\documentclass{article}
\usepackage[utf8]{inputenc}
\usepackage[english]{babel}
\usepackage{graphicx}
\usepackage{subfiles}
\usepackage{amsmath}
\usepackage{mathtools}

\usepackage[margin=1in]{geometry}
\usepackage{listings}
\usepackage{adjustbox}
\usepackage{multirow}
\usepackage{pdfpages}
\usepackage[font=small,labelfont=bf]{caption} 
\usepackage[nottoc]{tocbibind}
\usepackage[linesnumbered,lined,boxed,commentsnumbered]{algorithm2e}
\usepackage[width=.88\textwidth]{caption}

\author{Emily Morris\qquad
        Kevin He\qquad 
        Yanming Li\qquad
        Yi Li\qquad 
        Jian Kang\\University of Michigan, Ann Arbor, MI 48109 }
\title{\pkg{SurvBoost}: An \proglang{R} Package for High-Dimensional Variable Selection in the Stratified Proportional Hazards Model via Gradient Boosting}

 \newcommand{\pkg}[1]{\textbf{#1}}
\newcommand{\code}[1]{\texttt{#1}}
\let\proglang=\textsf

\begin{document}

\lstloadlanguages{R}

\maketitle
 \thispagestyle{empty}
\abstract{
High-dimensional variable selection in the proportional hazards (PH) model has many successful applications in different areas. In practice, data may involve confounding variables that do not satisfy the PH assumption, in which case the stratified proportional hazards (SPH) model can be adopted to control the confounding effects by stratification of the confounding variable, without directly modeling the confounding effects. However, there is lack of computationally efficient statistical software for high-dimensional variable selection in the SPH model. In this work, an \proglang{R} package, \pkg{SurvBoost}, is developed to implement the gradient boosting algorithm for fitting the SPH model with high-dimensional covariate variables and other confounders. Extensive simulation studies demonstrate that in many scenarios \pkg{SurvBoost} can achieve a better selection accuracy and reduce computational time substantially compared to the existing \proglang{R} package that implements boosting algorithms without stratification. The proposed \proglang{R} package is also illustrated by an analysis of the gene expression data with survival outcome in The Cancer Genome Atlas (TCGA) study. In addition, a detailed hands-on tutorial for \pkg{SurvBoost} is provided. 
}

\section{Introduction}

Variable selection for high-dimensional survival data has become increasingly important in a variety of research areas. One of the most popular methods is based on the proportional hazards (PH) model. Many penalized regression methods including adaptive lasso and elastic net have been proposed for the PH model [\cite{lasso, regularizationPaths, penalizedCoxPH}]. Alternatively, boosting described by Buhlmann and Yu [~\cite{WICS:WICS55}] has been adopted for variable selection in regression models and the PH model via gradient descent techniques. It can have a better variable selection accuracy compared with other methods in many scenarios.  The \proglang{R} package \pkg{mboost} has been developed and become a powerful tool for variable selection and parameter estimation in complex parametric and nonparametric models via the boosting methods [\cite{boosting2010}]. It has been widely used in many applications. 

However, in many biomedical studies, the collected data may involve confounding variables that do not satisfy the PH assumption. For example, in cancer research you may argue that gender effects are not proportional, but we are more interested in selecting genes as the important risk factors for cancer survival. The PH assumption can reasonably be imposed on modeling the gene effects but not for gender effects.  In this case the stratified proportional hazards (SPH) models are needed. In particular, the data are often grouped into multiple strata according to confounding variables. The SPH model adjusts those confounding effects by fitting the Cox regression with different baseline hazards for different strata, while still assuming that the covariate effects of interest are the same across different strata and satisfy the proportional hazard assumption. 

The SPH model has a wide range of applications for survival analysis, but no computationally efficient statistical software are available for high-dimensional variable selection in the SPH model. To fill this gap, we  develop an \proglang{R} package, \pkg{SurvBoost},  to implement the gradient boosting algorithm for fitting the SPH model with high-dimensional covariates with adjusting confounding variables. \pkg{SurvBoost} implements the gradient decent algorithm for fitting both PH and SPH model. The algorithm for the PH model has been used for the additive Cox model in \pkg{mboost} package which cannot fit the SPH model to perform variable selection. In our \pkg{SurvBoost} package, we optimize the implementations which can reduce 30\%--50\% computational time. Additional options are available in the \pkg{SurvBoost} package to determine an appropriate stopping criteria for the algorithm. Another useful function assists in selecting stratification variables, which may improve model fitting results.

The rest of the paper is organized as follows: In Section 2, we will provide a brief overview of the gradient boosting method for the SPH model along with the algorithm stopping criteria. In Section 3, we show that \pkg{SurvBoost} can achieve a better selection accuracy and reduce computational time substantially compared with \pkg{mboost}. In Section 4, we provide a detailed hands-on tutorial for \pkg{SurvBoost}.  In Section 5, we illustrate the proposed \proglang{R} package on an analysis of the gene expression data with survival outcome in The Cancer Genome Atlas (TCGA) study.  


\section{Methods}

\subsection{Stratified Proportional Hazards Model}

The Cox proportional hazards model is effective for modeling survival outcomes in many applications. An important assumption underlying this model is a constant hazard ratio, meaning that the hazard for one individual is proportional to that of any other individual. This is a strong assumption for many applications. Thus, one useful adaptation to this model is relaxing the strict proportional hazards assumption; one approach is to allow the baseline hazard to differ by group across the observations. This is known as the stratified proportional hazards (SPH) model.

Suppose the dataset consists of $n$ subjects. For $i = 1,\ldots, n$, denote by $T_i$ the observed time of event or censoring for subject $i$ and $\delta_{i}$ indicates whether or not an event occurred for subject $i$. Denote by $G$ the total number of strata and by $n_{g}$  the number of subjects in stratum $g$. Let  $g_{i}$ be the strata indicator for subject $i$. Suppose there are $p$ potential covariate variables of our interest to select. For $j = 1,\ldots, p$, let $x_{ij}$ be the covariate $j$ for subject $i$. For stratum $g = 1,\ldots, G$, the hazard of subject $i$ at time $t$  in stratum $g_i$ becomes 
$$ h(t, X_{i}, g_{i}) = \sum_{g=1}^{G} I_{[g_{i}=g]} \: h_{0g}(t) \: \exp\left\{ \sum_{j=1}^{p} \beta_{j} X_{ij} \right\},$$
where $I_{\mathcal{A}}$ is an event indicator where $I_{\mathcal{A}} = 1$ if $\mathcal{A}$ occurs and $I_{\mathcal{A}} = 0$ otherwise. The function $h_{0g}(t)$ represents the baseline hazard for group $g$. The coefficient $\beta_j$ represents the effect of covariate $j$. 
Allowing the baseline hazard to differ across strata allows flexibility often desired when proportional hazards is too strong. The SPH model can control effects of confounding variables through this stratification. The estimates of the effect of covariates remain constant across strata, so the model is still interpretable across all subjects.

\subsection{Gradient Boosting for SPH}

The log partial likelihood of the SPH model is 
$$ \ell(\beta) =  \sum_{i=1}^{n}\sum_{g=1}^{G}I_{[g_i = g]}\delta_{i} \: \left\{ X_{i}^{\top}\beta - \log\left(\sum_{\ell \in R_{ig}} \exp\{ X_{\ell }^{\top}\beta \} \right)\right\},$$
where $\beta = (\beta_1,\ldots,\beta_p)^{\top}$, $X_{i} = (X_{i1},\ldots,X_{ip})^{\top}$ and $R_{ig} = \{ \ell:T_{\ell} \geq T_{i},g_l = g \}$ for all $i$ with $g_i = g$ representing the set of at risk subjects in group $g$. We adopt the following gradient boosting algorithm to find the maximum  partial  likelihood estimate (MPLE). Let $S_{kg}(i,j) = \sum_{\ell \in R_{ig}} X_{\ell j}^k \exp\{ X_{\ell g}^{\top} \beta \}$ for $k = 0,1,2$.

\begin{minipage}{0.85\textwidth}
\begin{algorithm}[H]
\caption{Boosting gradient descent algorithm}
    \SetAlgoLined
    \KwData{$\{T_i,\delta_i, g_i,X_i\}_{i=1}^n$; Number of iterations $M$; Updating rate $\upsilon$}
    \KwResult{$\beta $.}
    \Begin{
    Initialize $\beta_{j} = 0 \: (j = 1,\ldots, p)$. \\
    \For{$m = 1,\ldots,M$}{
    \For{$j = 1,\ldots, p$}{
       Compute the first partial derivative with respect to $j$:
       $L_{1}(j) = \sum_{i=1}^{n} \sum_{g=1}^{G} I_{[g_i = g]} \delta_{i} \{X_{ij} - S_{1g}(i,j)/S_{0g}(i,i) \}$. 
     }
     
     Find  $j^* = \mathrm{argmax}_{j} \: |L_{1}(j)|$.
     
     Calculate the second partial derivative with respect to $j^{*}$:
     $L_{2}(j^*) =   \sum_{i=1}^{n}\sum_{g=1}^{G}I_{[g_i = g]} \delta_{i}\left[ \dfrac{S_{2g}(i,i)}{S_{0g}(i,i)}  - \left\{\dfrac{S_{1g}(i,j^{*})}{S_{0{g}}(i,i)}\right\}^{2} \right]$
        
    Update $\beta_{j^{*}} = \beta_{j^{*}} + \upsilon L_2(j^*)^{-1} L_1(j^*)$
    }
    }
    
\end{algorithm}
\end{minipage}

This algorithm updates variables one at a time, by selecting the variable which maximizes the first partial derivative. The number of iterations is important for ensuring a sufficient number of updates to the $\beta$ estimates, in addition to selecting the true signals [\cite{Kevin2016}].

\subsection{Stopping Criteria}

Selection of the number of boosting iterations is important. Over-fitting can occur if the number of iterations is too large  [\cite{jiang2004}]. The algorithm is less sensitive to the step size [\cite{boosting2007}]. 

\pkg{SurvBoost} provides several options for optimizing the number of iterations including: $k$-fold cross validation, Bayesian information criteria, change in likelihood, or specifying the number of variables to select. 

The Bayesian Information Criteria (BIC) is one approach for selecting the optimal number of boosting iterations. 

\begin{equation}
BIC = -2 \: \{l_{j}(\hat{\theta}_{j}) - l_{0}(\hat{\theta}_{0})\} + (p_{j} - p_{0}) \: \log(d),
\end{equation}
where $l_{j}(\hat{\theta}_{j})$ is the maximized likelihood for a model with $p_{j}$ selected variables and $l_{0}(\hat{\theta}_{0})$ is the maximized likelihood for the reference model with $p_{0}$ selected variables. The number of uncensored events is $d$. \cite{BIC} argue that replacing the sample size, $n$, with $d$ in the BIC calculation has better properties when dealing with censored survival models. 

The extended BIC is also useful in high dimensional cases; this approach penalizes for greater complexity 

\begin{equation}
EBIC = -2 \: l_{j}(\hat{\theta}_{j})  + p_{j}  \: \log(d) + 2  \: \gamma  \: \log  \: {{p}\choose{p_{j}}},
\end{equation}
where ${{p}\choose{p_{j}}}$ is the size of the class of models that model $j$ belongs to, $p$ is the total number of variables. The value of $\gamma$ is fixed between 0 and 1, selected to penalize at the appropriate rate. Selecting 0 will reduce this to the standard BIC. 

Cross validation is another approach which may be used to determine the stopping point. The goodness of fit function is calculated as suggested by \cite{regularizationPaths}. It is the log-partial likelihood of all the data using the optimal $\beta$ determined with data excluding fold $k$ ($\beta_{-k}$) minus the log-partial likelihood excluding fold $k$ ($\ell_{-k}$) of the data with the same $\beta$.

\begin{equation}
CV_{k}(m) = -[\ell\{\beta_{-k}(m)\} - \ell_{-k}\{\beta_{-k}(m)\}],\\
\end{equation}

Where $m$ is the current number of iterations and $k$ indicates the subset of data being excluded. 

Change in likelihood is another approach incorporated in the package. This method stops iterating once a small change in likelihood, specified in the function, is reached. 

\begin{equation}
\Delta \ell = -\left[\ell(\beta(m)) - \ell(\beta(m+1))\right] < \alpha,\\
\end{equation}

Where $\alpha$ is a small constant. Default change in likelihood, used in simulations, is a change of 0.001.

\section{Simulation Studies}

This section compares the variable selection performance to a competing \proglang{R} package, \pkg{mboost} [\cite{boostingTutorial}]. 

\paragraph{Stratified Data} Stratified data was simulated such that censoring rates were relatively constant across groups and the expected survival time differed by group. These assumptions mimic realistic settings such as those encountered with data grouped by hospital or facility. 

For this simulation 1,000 observations were generated into ten strata; each strata had a different baseline hazard following a Weibull distribution. The Weibull distribution shape parameter was 3 for all strata, and the scale parameter varied across strata from $e^{-1}$ to $e^{-15}$ with ten evenly spaced intervals. There were 100 true signals among 4,000 variables with true magnitude of 2 or -2. There was uniform censoring from time 0 to 200. Ten of these data sets were generated. 

The following example demonstrates the importance of the stopping criteria. \pkg{SurvBoost} has five options for specifying the number of iterations as described in section 2.2. Selecting an appropriate number of iterations depends on the goals of the analysis. For example, if the goal is to achieve high sensitivity cross validation or extended BIC may be the best approach.

This simulation presents the performance of \pkg{SurvBoost} compared to the \proglang{R} package \pkg{mboost}. The boosting algorithm implemented in \pkg{mboost} is very similar to that of \pkg{SurvBoost} but does not allow stratification. With K-fold cross validation incorporated in \pkg{mboost}, we will compare results using cross validation and specifying a fixed number of iterations. The other stopping methods are not available in \pkg{mboost}. The performance can be compared by measures such as sensitivity and mean squared error. Table 1 presents the results of ten simulated data sets, comparing the boosting algorithm using several different stopping procedures to both default settings and cross validation methods of the package \pkg{mboost}. In this simulation, \pkg{mboost} selects fewer variables on average resulting in fewer false positives and more false negatives. Additionally the mean squared error is higher than that of all the \pkg{SurvBoost} options. 

Runtime is also an important factor with this algorithm. Stratification speeds up the algorithm as seen in the first simulation. All runtimes were generated on a MacBook with 2.9GHz Intel Core i5 and 16GB memory.

 \begin{table}[ht]
\centering
\small
\begin{tabular}{lrrrrrrrr}
  \hline
 & stopping &number   & \multirow{2}{*}{Se} & \multirow{2}{*}{Sp} & \multirow{2}{*}{FDR} & \multirow{2}{*}{MSE} & number of & runtime \\ 
& method&selected & & & & & iterations & (seconds)\\  
  \hline
  \hline
SurvBoost & \multirow{2}{*}{fixed} & 110 (2) & 0.92 (.02) & 1.00 (.00) & 0.16 (.02) & 380 (1) & 500 (0) & 44 (2) \\ 
    mboost & & 94 (5) & 0.78 (.03) & 1.00 (.00) & 0.17 (.03) & 387 (1) & 500 (0) & 24 (1) \\ 
\hline
  SurvBoost & \multirow{2}{*}{cv} & 214 (13) & 1.00 (.00) & 0.97 (.00) & 0.53 (.03) & 297 (1) & 5000 (0) & 2601 (82) \\ 
    mboost & &  275 (8) & 1.00 (.00) & 0.96 (.00) & 0.64 (.01) & 333 (1) & 5000 (1) & 2942 (95) \\ 
    \hline
\hline
  SurvBoost & \# selected &  100 (0) & 0.85 (.03) & 1.00 (.00) & 0.15 (.03) & 384 (1) & 381 (29) & 36 (2)  \\ 
  SurvBoost & likelihood &  118 (2) & 0.96 (.01) & 0.99 (.00) & 0.18 (.02) & 375 (1) & 633 (29) & 67 (4) \\ 
 SurvBoost & EBIC & 126 (5) & 0.99 (.03) & 0.99 (.00) & 0.21 (.03) & 365 (1) & 998 (3) & 173 (4) \\
   \hline
\end{tabular}
\caption {Results from simulation with approximately 1,500 observations in 10 strata and 4,000 variables to be selected. The table presents averages with the standard deviation, in parentheses, from ten simulated datasets. Sensitivity (Se) is calculated as the proportion of true positives out of the total number of true signals. Specificity (Sp) is calculated as the proportion of true negatives out of the total number of variables that are not true signals. }
\end{table}

\paragraph{Unstratified Data} Another simulation was used to compare performance of our method to \pkg{mboost} when stratification is not necessary for appropriate modeling. In this case one thousand observations were generated without stratification. The baseline hazard followed a Weibull distribution, with shape parameter equal to 3 and scale equal to 2. The true $\beta$ contained 100 true signals of magnitude 2 or -2 out of 1,000 variables. 

We can observe in Table 2 that \pkg{SurvBoost} performs similarly to \pkg{mboost} under these conditions. \pkg{mboost} tends to select fewer variables than \pkg{SurvBoost}, so in this simulation \pkg{mboost} has fewer false positives and more false negatives compared to \pkg{SurvBoost}.

\begin{table}[ht]
\centering
\small
\begin{tabular}{lrrrrrrrr}
  \hline
 & stopping &number  & \multirow{2}{*}{Se} & \multirow{2}{*}{Sp} & \multirow{2}{*}{FDR} & \multirow{2}{*}{MSE} & number of & runtime \\ 
& method&selected & &  &  & & iterations & (seconds)\\  
  \hline
  \hline
SurvBoost & \multirow{2}{*}{fixed} & 104 (5) & 0.78 (.02) & 0.97 (.01) & 0.25 (.04) & 379 (1) & 500 (0) & 4 (1)  \\ 
    mboost & & 82 (5) & 0.64 (.02) & 0.98 (.00) & 0.22 (.04) & 387 (0) & 500 (0) & 10 (0)  \\ 
\hline
  SurvBoost & \multirow{2}{*}{cv} & 213 (13) & 1.00 (.00) & 0.87 (.01) & 0.53 (.03) & 299 (2) & 5000 (0) & 391 (24) \\ 
    mboost &   & 181 (13) & 1.00 (.01) & 0.91 (.01) & 0.45 (.04) & 333 (2) & 5000 (1) & 1222 (44) \\ 
    \hline
\hline
  SurvBoost & \# selected &  100 (0) & 0.76 (.03) & 0.97 (.00) & 0.24 (.03) & 380 (2) & 453 (42) &  4 (0) \\ 
  SurvBoost & likelihood & 108 (5) & 0.81 (.03) & 0.97 (.01) & 0.25 (.03) & 377 (1) & 549 (18)  & 6 (0) \\ 
 SurvBoost & EBIC & 38 (1) & 0.29 (.01) & 0.99 (.00) & 0.09 (.25) & 389 (0) & 300 (2) & 13 (0) \\
   \hline
\end{tabular}
\caption {Results from simulation with approximately 1,000 observations and 1,000 variables to be selected. The table presents averages with the standard deviation from ten simulated datasets.}
\end{table}

\section{Illustration of Package}

This section provides a brief tutorial on how to use this package based on simulated data. In order to install the package, several other \proglang{R} packages must be installed. The code relies on \pkg{Rcpp}, \pkg{RcppArmadillo}, and \pkg{RcppParallel} in order to improve computational speed. Additionally the \pkg{survival} package is used for simulation and post selection inference and will be required for installation of \pkg{SurvBoost}. The following line of \proglang{R} code installs the package.

\begin{lstlisting}[language=R, belowskip=-0.8 \baselineskip]
install.packages("SurvBoost_0.1.0.tar.gz",type="source",repos = NULL)
\end{lstlisting}

\subsection{Model fitting} The \code{boosting\_core()} function requires similar inputs to the familiar \code{coxph()} function from the package \pkg{survival}.
\code{boosting\_core(formula, data = matrix(), rate = 0.01,control = 500, ...)}
The input \code{formula} has the form \code{Surv(time, death)} {\raise.17ex\hbox{$\scriptstyle\mathtt{\sim}$}}
 \code{variable1 + variable2}. The input \code{data} is in matrix form or a data frame. Two additional parameters must be specified for the boosting algorithm: \code{rate} and \code{control}. \code{Rate} is the step size in the algorithm, although choice of this may not impact the performance too significantly [\cite{boosting2007}], default value is set to 0.01. Selecting an appropriate number of iterations to run the algorithm will, however, have a greater impact on the results. The last input \code{control} is used to determine the number of iterations to run the algorithm, default value is 500.

\begin{table}[ht]
\centering
\begin{tabular}{ll}
  \hline
Call & Method \\
\hline
\hline
boosting\_core(formula, data) & fixed mstop = 500 \\
boosting\_core(formula, data, control=1000) & fixed mstop = specified value \\
boosting\_core(formula, data, control\_method="cv") & 10-fold cross validation \\
boosting\_core(formula, data, control\_method="num\_selected", & number selected, need to specify \\ \hspace{25mm}control\_parameter = 5) & number of variables \\
boosting\_core(formula, data, control\_method="likelihood") & change in likelihood \\
boosting\_core(formula, data, control\_method="BIC") & minimum BIC or EBIC \\
boosting\_core(formula, data, control\_method="AIC") & minimum AIC \\

   \hline
\end{tabular}
\caption {Stopping criteria options for boosting\_core function.}
\end{table}

\begin{table}[ht]
\centering
\begin{tabular}{ll}
  \hline
Function & Result \\
\hline
\hline
summary.boosting() & prints summary of variable selection and estimation \\
modelfit.boosting() & prints summary of model and data \\
plot.boosting() & plots variable selection frequency \\
predict.boosting() & generates predicted hazard ratio for each observation or a new dataset\\
   \hline
\end{tabular}
\caption {Functions available in \pkg{SurvBoost} package. Every function accepts a boosting object input to generate the corresponding result. }
\end{table}

\subsection{Simple example}

We present a simple example demonstrating the convenience of using the package for stratified data. We simulate survival data for five facilities with different constant baseline hazards.  
\begin{lstlisting}[language=R, belowskip=-0.8 \baselineskip]
R > TrueBeta
 [1] 0.5 0.5 0.0 0.0 0.0 -0.5 0.5 0.5 0.0 0.0
R > set.seed(123)
R > data_small <- simulate_survival_cox(true_beta=TrueBeta,
               base_hazard="auto",
               num_facility=5,
               input_facility_size=100, cov_structure="ar",
               block_size=5, rho=0.6, censor_dist="unif",
               censor_const=2, tau=Inf, normalized=F)

\end{lstlisting}

We have $p = 10$ and $|\beta_j|$ ranges from 0 to 0.5. There are five ``facilities'' with average size of 100, and $n$ is approximately 500. The covariance structure within the blocks is AR(1) with correlation 0.6. 
The censoring rate is about 33\%. In this case the variable \textit{facility\_idx} indicates the variable to stratify on in the survival model; each ``facility'' in this simulated data has a different baseline hazard function. 

Another feature of the package assists with determining variables to stratify on if this information is unknown. The function \textit{strata.boosting} will print box plots and a summary table of the survival time grouped by splits in a the specified variable. The variable can be categorical or continuous; if continuous, the function will split on the median value to demonstrate whether there appears to be a difference in the survival time distribution for the two groups.

\begin{lstlisting}[language=R]
R > strata.boosting(data_small$facility_idx, data_small$time)

  as.factor(x)          Min        Q1    Median       Q3      Max
1            1 0.0046772744 0.1163388 0.3108169 1.096236 1.693283
2            2 0.0005600448 0.1422992 0.5849665 1.270754 1.951286
3            3 0.0057943145 0.1371938 0.9125127 1.314191 1.989180
4            4 0.0042511208 0.1998902 0.5797646 1.437124 1.960646
5            5 0.0015349222 0.1283325 0.5896426 1.325094 1.873137
\end{lstlisting}
\includegraphics[width=16cm]{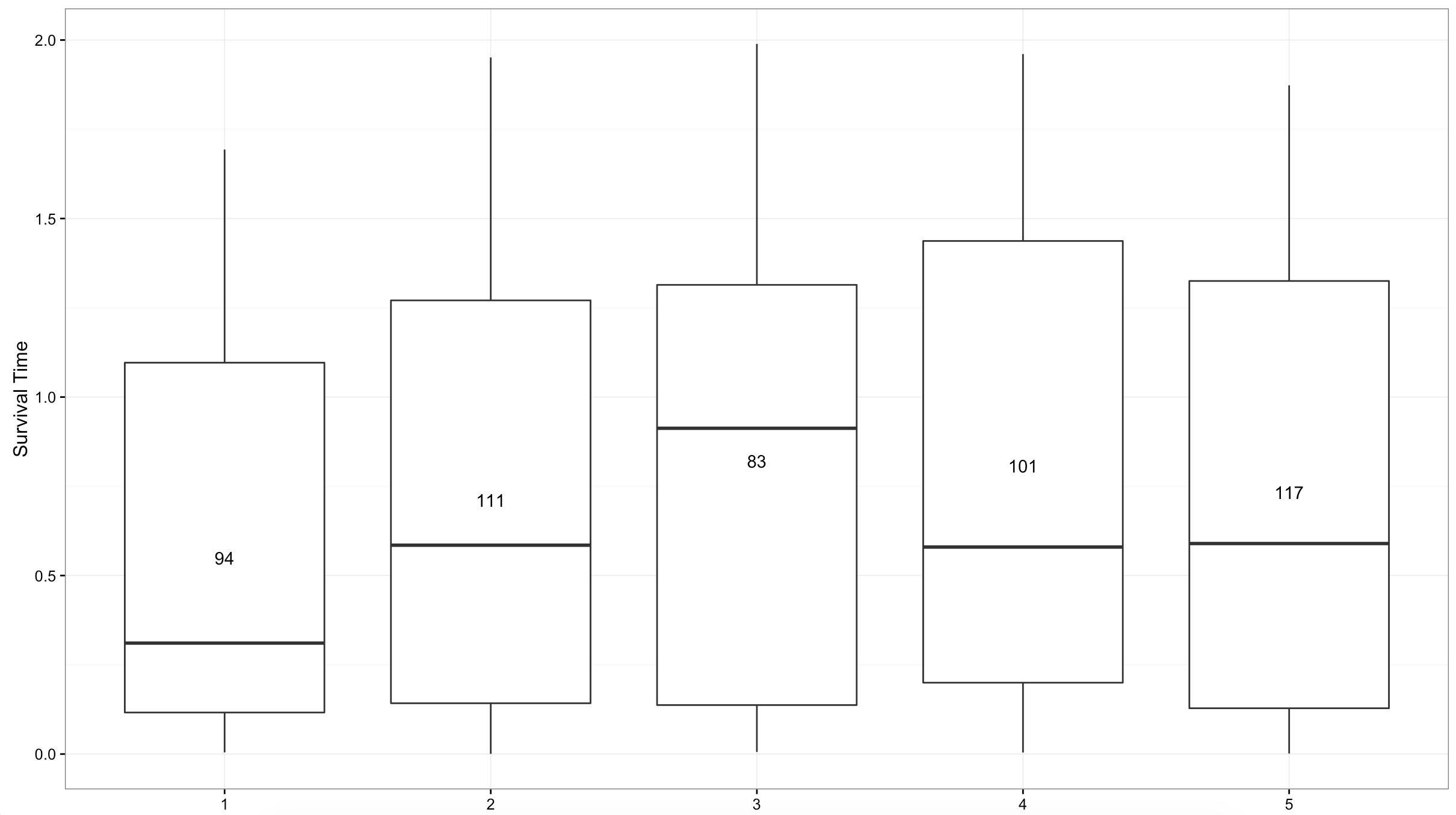}
\captionof{figure}{Box plots of survival time by facility index in simulated data generated by the function \textit{strata.boosting}.}

\vspace{0.5cm}

Simulated data includes a vector of survival or censoring time, \textit{time}, indicator of an event, \textit{delta}, and matrix of covariates, \textit{Z}. Then generate the formula including all possible variables for selection. 
\begin{lstlisting}[language=R]
R > time <- data_small$time
R > delta=data_small$delta
R > Z <- as.matrix(data_small[,-c(1,2,3)])

R > covariates <- paste("strata(facility_idx)+", paste(colnames(Z), 
    collapse = "+"))
R > formula <- as.formula(paste("Surv(time,delta)~", covariates))
\end{lstlisting}

Run the \code{boosting\_core()} function to obtain the variables selected. This example uses the number of iterations control as a fixed input of 75 and update rate of 0.1. 

\begin{lstlisting}[language=R]
R > test1 <-  boosting_core(formula,    
+               data=data_small,    
+               rate=0.1,     
+               control=75)               
R > summary.boosting(test1)

Surv(time, delta) ~ V1 + V2 + V6 + V7 + V8 + strata(strata)

Coefficients:
       V1        V2        V6        V7        V8 
0.5276104 0.3898193 -0.4355044 0.4469272 0.4309359 

 Number of iterations:  75
\end{lstlisting}
Function \code{summary.boosting()} displays the variables which are selected as well as the coefficient estimates and the number of boosting iterations performed. Set the argument \textit{all\_beta} = TRUE to see all the variables, not just those selected. More detailed information about the model can be obtained through the function \code{modelfit.boosting()}.

\begin{lstlisting}[language=R]
R > modelfit.boosting(test1)
Call:
 data:  
 n =  506 
 Number of events =  346 
 Number of boosting iterations: mstop =  75 
 Step size =  0.1

 Coefficients: 
       V1        V2        V6        V7        V8 
0.5276104 0.3898193 -0.4355044 0.4469272 0.4309359 
\end{lstlisting}

To use a different method for the number of boosting iterations use the arguments \textit{control\_method} and \textit{control\_parameter}. For example,
\begin{lstlisting}[language=R]
R > test2 <- boosting_core(formula, data=data_small, rate=0.1,
    control_method="num_selected", control_parameter=5)
R > summary.boosting(test2)
Surv(time, delta) ~ V1 + V2 + V6 + V7 + V8 + strata(strata)

Coefficients:
        V1         V2         V6         V7         V8 
0.11828718 0.11021464 -0.05292158 0.25561965 0.05199151 

 Number of iterations:  10
 \end{lstlisting}
 
 This option iterates until the specified number of variables, 5 in this example, are selected. See methods for other stopping criteria. \\

The \code{plot.boosting()} function displays a plot of the selection frequency by the number of iterations. Another option of the \code{plot.boosting()} function is to plot the coefficient paths of each variable by the number of boosting iterations. See Figures 2 and 3. 

\begin{center}
\includegraphics[width=10cm]{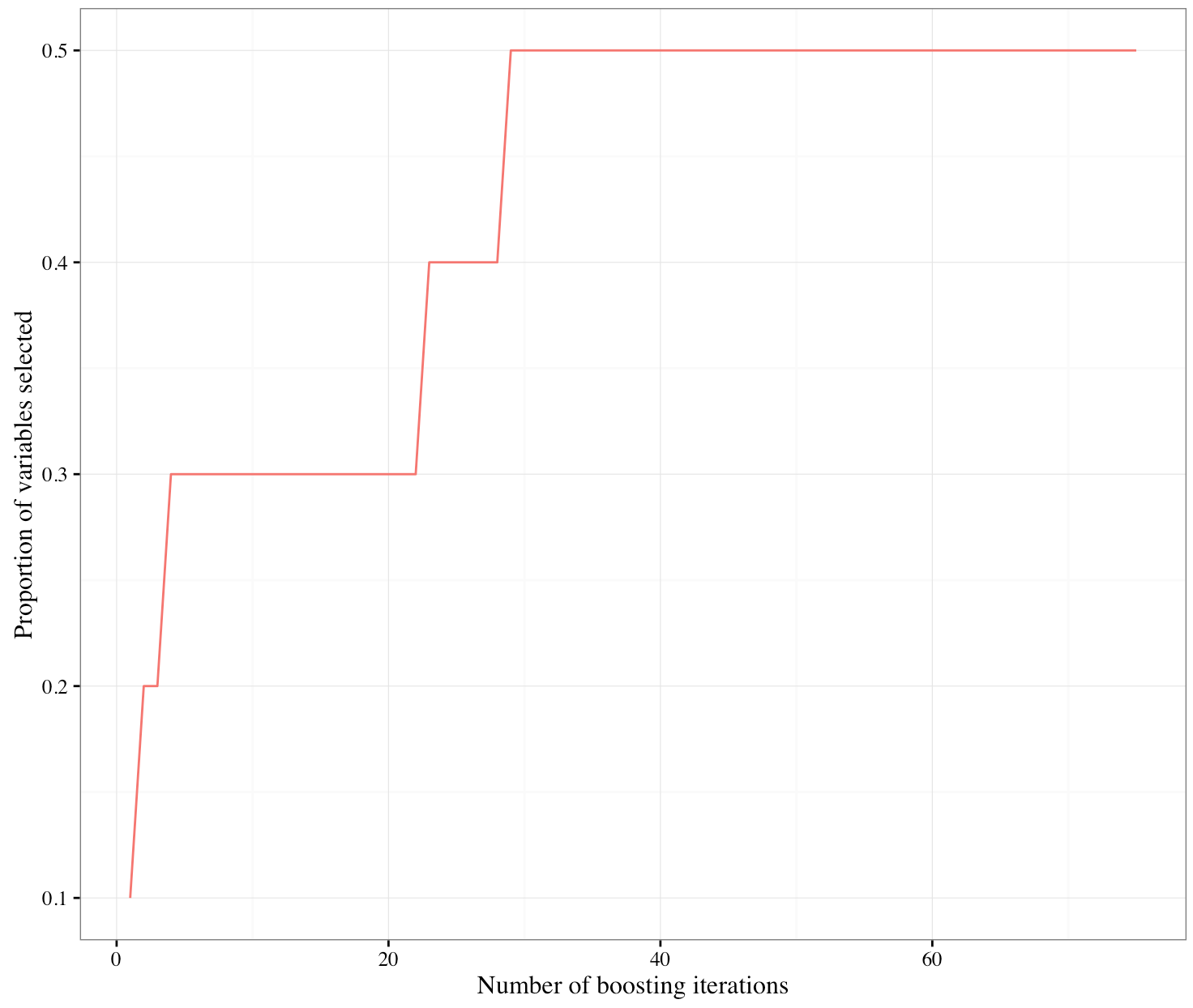}
\captionof{figure}{Plot generated by plot.boosting function, variable selection frequency by number of boosting iterations.}
\end{center}

\begin{center}
\includegraphics[width=10cm]{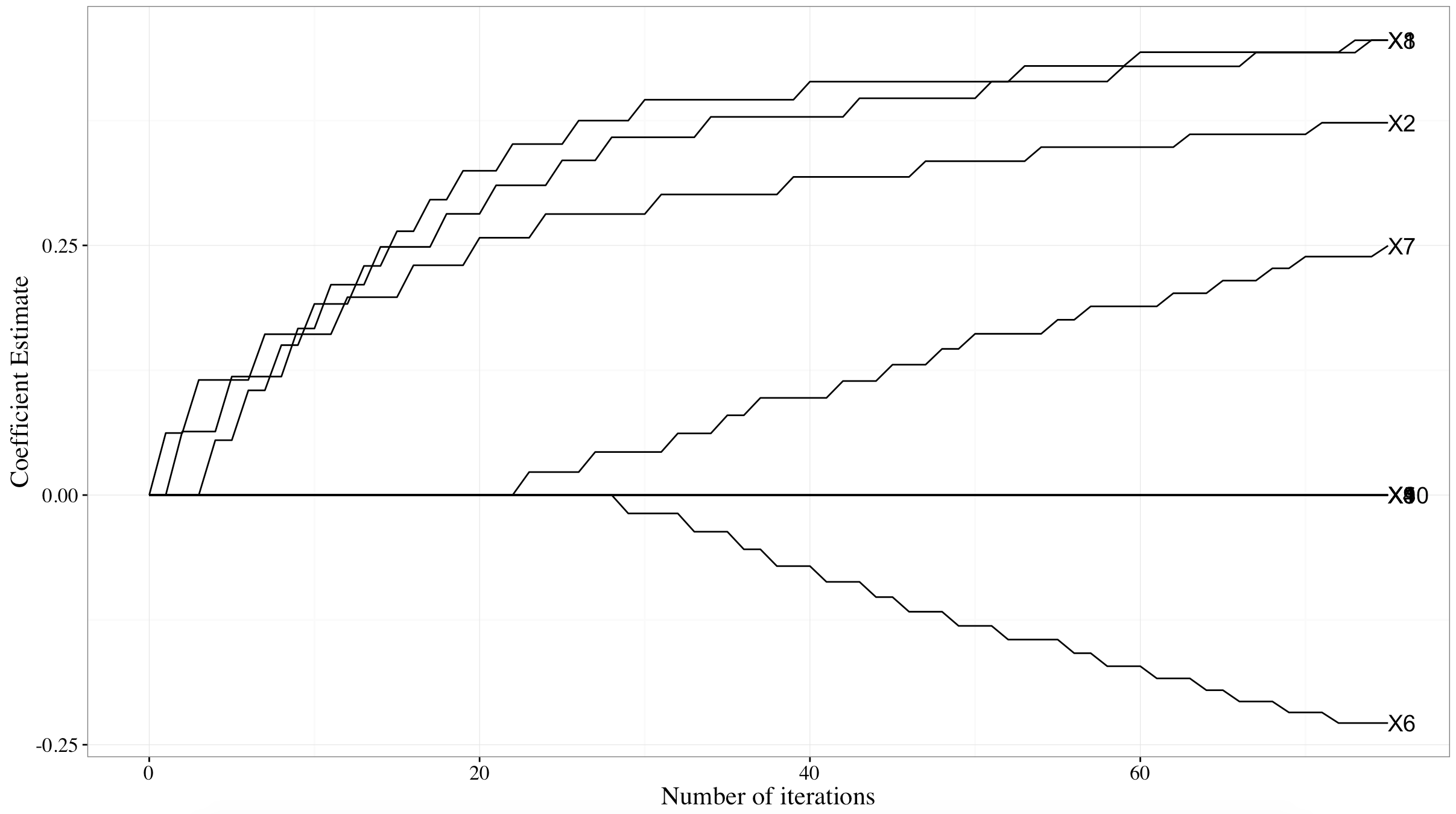}
\captionof{figure}{Plot generated by plot.boosting function with option ``coefficients'', coefficient paths for variables selected by number of boosting iterations.}
\end{center}

\vspace{0.5cm}
The function \code{predict.boosting()} provides an estimate of the hazard ratio for each observation in the dataset provided relative to the average of $p$ predictors. 

\begin{lstlisting}[language=R]
R > predict.boosting(test1)[1:6]
46.385476  1.823920 42.049932 16.427860  4.013200  2.243711
\end{lstlisting}

The model selected using boosting can be refit with \code{coxph()} for post selection inference. The function \code{inference.boosting()} will perform this refitting and output the coefficient estimates with corresponding standard errors and p-values. 

\begin{lstlisting}[language=R]
R > fmla <- summary.boosting(test1)$formula
R > inference.boosting(fmla, data=data_small)
Call:
coxph(formula = fmla, data = data)

  n= 506, number of events= 371 

       coef exp(coef) se(coef)      z Pr(>|z|)    
V1  0.59181   1.80726  0.07454  7.940 2.00e-15 ***
V2  0.48079   1.61736  0.06948  6.920 4.53e-12 ***
V6 -0.51830   0.59553  0.07145 -7.254 4.05e-13 ***
V7  0.51108   1.66709  0.08479  6.028 1.66e-09 ***
V8  0.54758   1.72907  0.07116  7.695 1.42e-14 ***
---
Signif. codes:  0 '***' 0.001 '**' 0.01 '*' 0.05 '.' 0.1 ' ' 1

   exp(coef) exp(-coef) lower .95 upper .95
V1    1.8073     0.5533    1.5616    2.0915
V2    1.6174     0.6183    1.4114    1.8533
V6    0.5955     1.6792    0.5177    0.6851
V7    1.6671     0.5998    1.4118    1.9685
V8    1.7291     0.5783    1.5040    1.9879

Concordance= 0.762  (se = 0.036 )
Rsquare= 0.487   (max possible= 0.997 )
Likelihood ratio test= 338.1  on 5 df,   p=0
Wald test            = 287.8  on 5 df,   p=0
Score (logrank) test = 299.1  on 5 df,   p=0
\end{lstlisting}

\section{TCGA Data Example}

Data from three breast cancer cohorts was used to demonstrate this method on data outside of the simulation framework. There were 578 patients included in the combined data, with 8864 variables measured for each patient: 8859 genes and 5 phenotypic variables. The phenotype variables included age at diagnosis, tumor size, cancer stage, progesterone-receptor status, and estrogen-receptor status. The data can be downloaded from The Cancer Genome Atlas (TCGA) [\cite{Caldas, Chin, Miller}].  

The patients were split into two cohorts depending on their cancer stage and tumor size. One cohort contained patients with the less severe prognosis, cancer stage of one and tumor size less than the median; the other cohort contained those with cancer stage greater than one and/or with a tumor larger than the median size. \\

\begin{lstlisting}[language=R]
R > fit.plot <- survfit(Surv(survival_time, survival_ind) ~ as.factor(severity), data=data)
R > ggsurvplot(fit.plot,
           conf.int = TRUE,
           risk.table = TRUE,
           risk.table.col="strata",
           ggtheme = theme_bw(), palette = "grey")
\end{lstlisting}

\includegraphics[width=16cm]{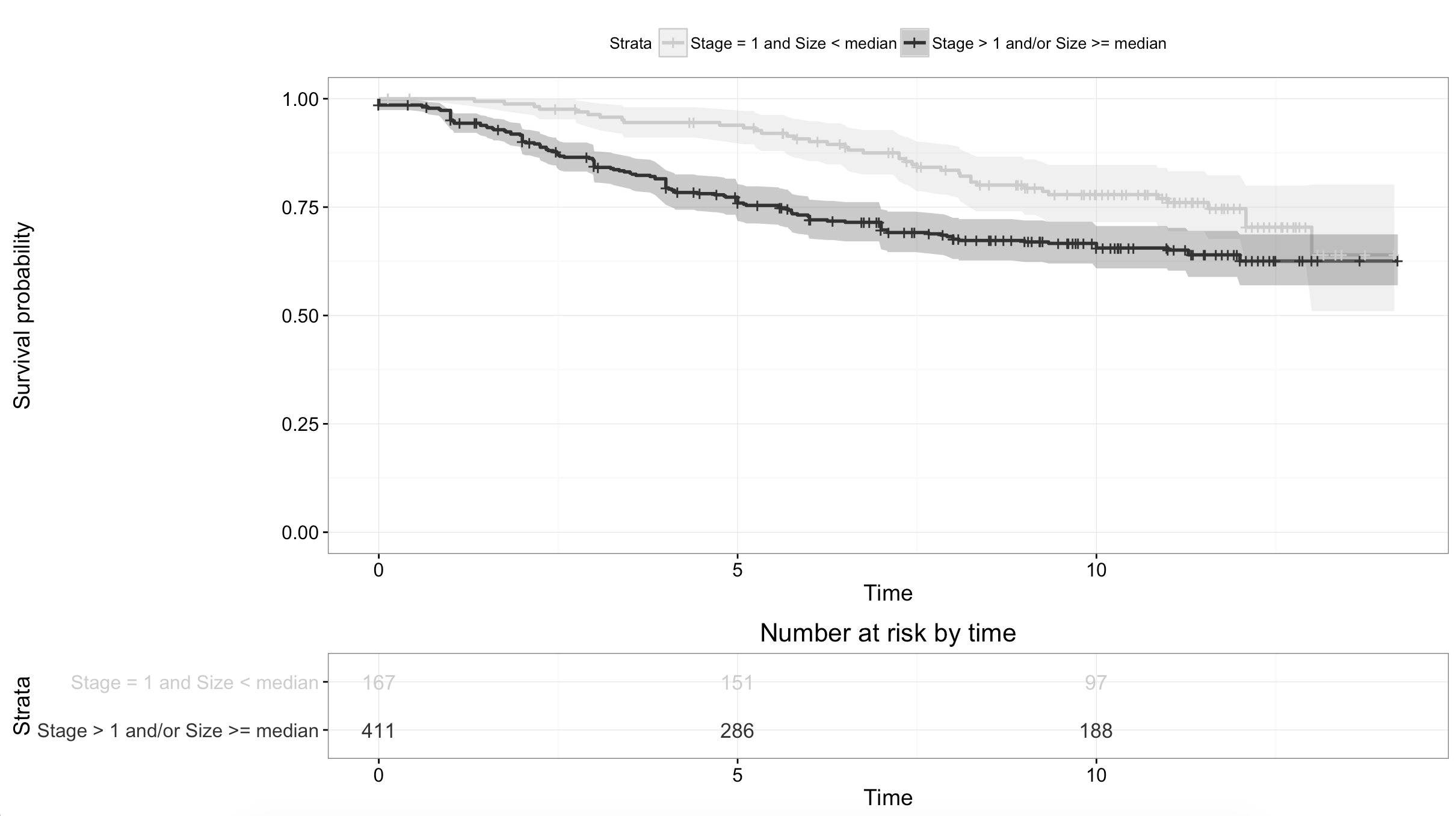}
\captionof{figure}{Survival curves for the two strata based on cancer stage and tumor size. }

This plot demonstrates that the proportional hazards assumption may not hold in this case. Stratifying based on this criteria generates the following results. \\

Using stability selection [\cite{Buhlmann2010}], 14 variables were identified with selection frequencies greater than 50\% from 50 iterations of subsampling. Age and progesterone-receptor status were selected in addition to 12 genes. The boosting algorithm was performed with the number of iterations fixed at the sample size of 578.

\begin{center}
\includegraphics[width=12cm]{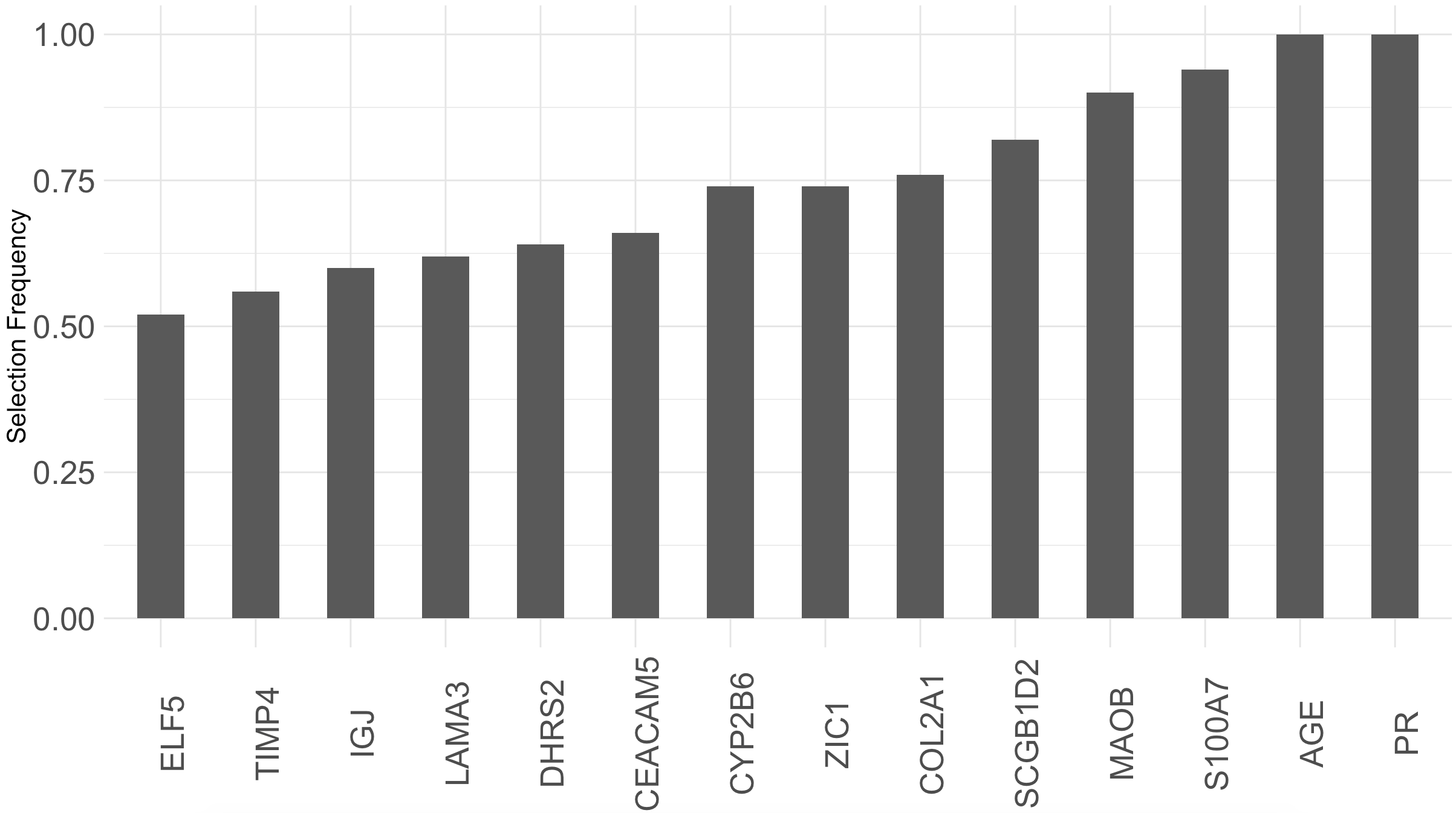}
\captionof{figure}{Selection frequencies for genes or phenotype variables that were selected at least 50\% of the time with stability selection.}
\end{center}

Several of the genes selected in this analysis have been previously identified as having an association with breast cancer. Psoriasin (S100A7) has been associated with breast cancer  [\cite{BC_gene1}]. Several studies have found COL2A1 to be part of gene signatures for predicting tumor recurrence [\cite{BC_gene2}, \cite{WANG2005671}]. Other genes selected that have been identified as part of a gene signature or association with breast cancer tumor progression risk include: ZIC1 [\cite{BC_geneZIC1}), CYP2B6 (\cite{BCgeneCYP2B6}], ELF5 [ \cite{BCgeneELF5}], IGJ [\cite{BC_geneZIC1}], DHRS2 [\cite{Krijgsman2012}], and CEACAM5 [\cite{Blumenthal2007}]. \pkg{Mboost} using the same criteria but without a stratified model only identifies one gene of importance, MC2R, demonstrating the utility of the SPH model in this context. 

\section{Conclusion}
In this article, we introduce a new \proglang{R} package \pkg{SurvBoost} which implements the gradient boosting algorithm for high-dimensional variable selection in the stratified proportional hazards (SPH) model, while most existing \proglang{R} packages, such as \pkg{mboost} only focus on the proportional hazards model. In the simulation studies, we show that \pkg{SurvBoost} can improve the model fitting and achieve better variable selection accuracy for the data with stratified structures. In addition, we optimize the implementations of the gradient boosting in both the SPH and the PH models. For the PH model fitting, \pkg{SurvBoost} can reduce about 30\%-50\% computational time compared to \pkg{mboost}. In the future, we plan to extend the package to handle more complex survival data such as left-truncation data and interval censoring data.



\bibliographystyle{plain}

\newpage
  Emily Morris\\
  Department of Biostatistics\\
  University of Michigan\\
  1415 Washington Heights, Ann Arbor MI 48109\\
  E-mail: emorrisl@umich.edu\\
  
  Jian Kang\\
  Department of Biostatistics\\
  University of Michigan\\
  1415 Washington Heights, Ann Arbor MI 48109\\
  E-mail: jiankang@umich.edu\\

\end{document}